\documentclass[twocolumn,superscriptaddress,showpacs,floatfix,aps,prl]{revtex4}
\usepackage{graphicx}
\usepackage{units}
\usepackage{bm}
\usepackage{amsmath}
\usepackage{amsfonts}
\usepackage{amssymb}
\usepackage[thinspace,squaren,pstricks,derivedinbase,derived]{SIunits}
\usepackage{CJKutf8}
\usepackage[pdfpagelabels, colorlinks, citecolor=blue]{hyperref}
\renewcommand{\vec}[1]{{\mathbf #1}}

\begin{document}

\title{Temperature Dependence of Magnetic Excitations: Terahertz Magnons above the Curie Temperature}

\author{H.J. Qin} \affiliation {Max-Planck-Institut f$\ddot{u}$r Mikrostrukturphysik, Weinberg 2, D-06120 Halle, Germany}
\author{Kh. Zakeri}
\email{khalil.zakeri@partner.kit.edu}
\affiliation {Heisenberg Spin-dynamics Group, Physikalisches Institut, Karlsruhe Institute of Technology, Wolfgang-Gaede-Str. 1, D-76131 Karlsruhe, Germany}\affiliation {Max-Planck-Institut f$\ddot{u}$r Mikrostrukturphysik, Weinberg 2, D-06120 Halle, Germany}
\author{A. Ernst} \affiliation {Max-Planck-Institut f$\ddot{u}$r Mikrostrukturphysik, Weinberg 2, D-06120 Halle, Germany}
\author{J. Kirschner}\affiliation {Max-Planck-Institut f$\ddot{u}$r Mikrostrukturphysik, Weinberg 2, D-06120 Halle, Germany}

\begin{abstract}
When an ordered spin system of a given dimensionality undergoes a second order phase transition the dependence of the order parameter i.e. magnetization on temperature can be well-described by thermal excitations of elementary collective spin excitations (magnons). However, the behavior of magnons themselves, as a function of temperature and across the transition temperature $T_C$, is an unknown issue. Utilizing spin-polarized high resolution electron energy loss spectroscopy we monitor the high-energy (terahertz) magnons, excited in an ultrathin ferromagnet, as a function of temperature. We show that the magnons' energy and lifetime decrease with temperature. The temperature-induced renormalization of the magnons' energy and lifetime depends on the wave vector. We provide quantitative results on the temperature-induced damping and discuss the possible mechanism e.g., multi-magnon scattering. A careful investigation of physical quantities determining the magnons' propagation indicates that terahertz magnons sustain their propagating character even at temperatures far above $T_C$.
\end{abstract}

\date{\today}
\maketitle

One of the most fundamental concepts in condensed-matter physics is the universal behavior of systems, undergoing a second-order phase transition \cite{Domb2000}. The phase transition is associated with a symmetry breaking, occurring at the critical temperature. The degree of broken symmetry is represented by an order parameter, which is a continuous function of temperature and is expected to exhibit a power law behavior. In an ordered spin system, e.g. a ferromagnetic solid, the dependence of the order parameter i.e. magnetization on the temperature is described by thermal excitations of elementary collective spin excitations (magnons). This description provides an excellent account for the universal behavior of the phase transition. Although the phase transition in spin systems has been investigated extensively, the behavior of magnons, themselves, as a function of temperature is poorly understood.

Since the beginning of the development of quantum theory of magnetism, one of the most important and fundamental questions has been: how do the magnons in a ferromagnet behave across and above the Curie temperature ($T_C$)? This longstanding question has still remained open, mainly due to the lack of full experimental magnon spectrum over a wide range of wave vector, energy and temperature.
Only a few experimental results, obtained by means of inelastic neutron scattering (INS) on bulk Fe and Ni have been reported at the earlier stage of the development of this technique \cite{Lynn1975, Mook1986, Wicksted1984, Wicksted1984a}. The results have led to major controversies as the experiments have been performed in the so-called constant energy scan mode which do not provide any information regarding the lifetime of excitations. In the case of low-dimensional itinerant magnets, e.g. magnetic thin films and nanostructures, the situation is even worse; no experimental data are available on magnetic excitations above $T_C$, since INS does not have the desired sensitivity for probing excitations in low-dimensional magnets.

Beside their importance in fundamental physics, magnons in low dimensional magnets are also important for the research in the field of magnonics, where the main idea is to use the ultrafast magnons for information processing. For this purpose only the magnons which can propagate a certain distance are useful. Here, the question of interest is how the propagation behavior of ultrafast terahertz magnons, generated in a device made of an ultrathin film, changes as a function of temperature. In particular, it is of great interest to know whether the well-defined terahertz excitations observed at low temperatures can survive at high temperatures, above $T_C$.

Spin caloritronics focuses on the manipulation of spins with heat currents \cite{Bauer2012}. One of the fascinating effects in this field is the so-called spin-dependent Seebeck effect, which refers to the generation of a spin voltage induced by a temperature gradient in a ferromagnet \cite{Bauer2012}. The effect has been attributed to collective spin excitations. Here a fundamental question is: can the spin-Seebeck effect exist in the paramagnetic state, where the long range magnetic order is absent? In order to answer this question one should precisely know the behavior of magnons as a function of temperature and above $T_C$.

In this Letter by investigating terahertz magnons in a model system i.e. an ultrathin ferromagnetic film grown on a nonmagnetic substrate we report, for the first time, on the temperature dependence of magnetic excitations across the transition temperature. We demonstrate that although the temperature effects lead to the renormalization of magnons' energy and lifetime, the excitations are still well-defined. Above $T_C$, terahertz magnons exhibit a finite lifetime and behave as propagating quasiparticles within their short propagation distance. The extrapolation of our experimental data to higher temperatures indicates that this propagating character can survive even at temperatures much above $T_C$.

\begin{figure}[t]
\vspace{12pt} \center
\resizebox*{0.99\columnwidth}{!}{\includegraphics{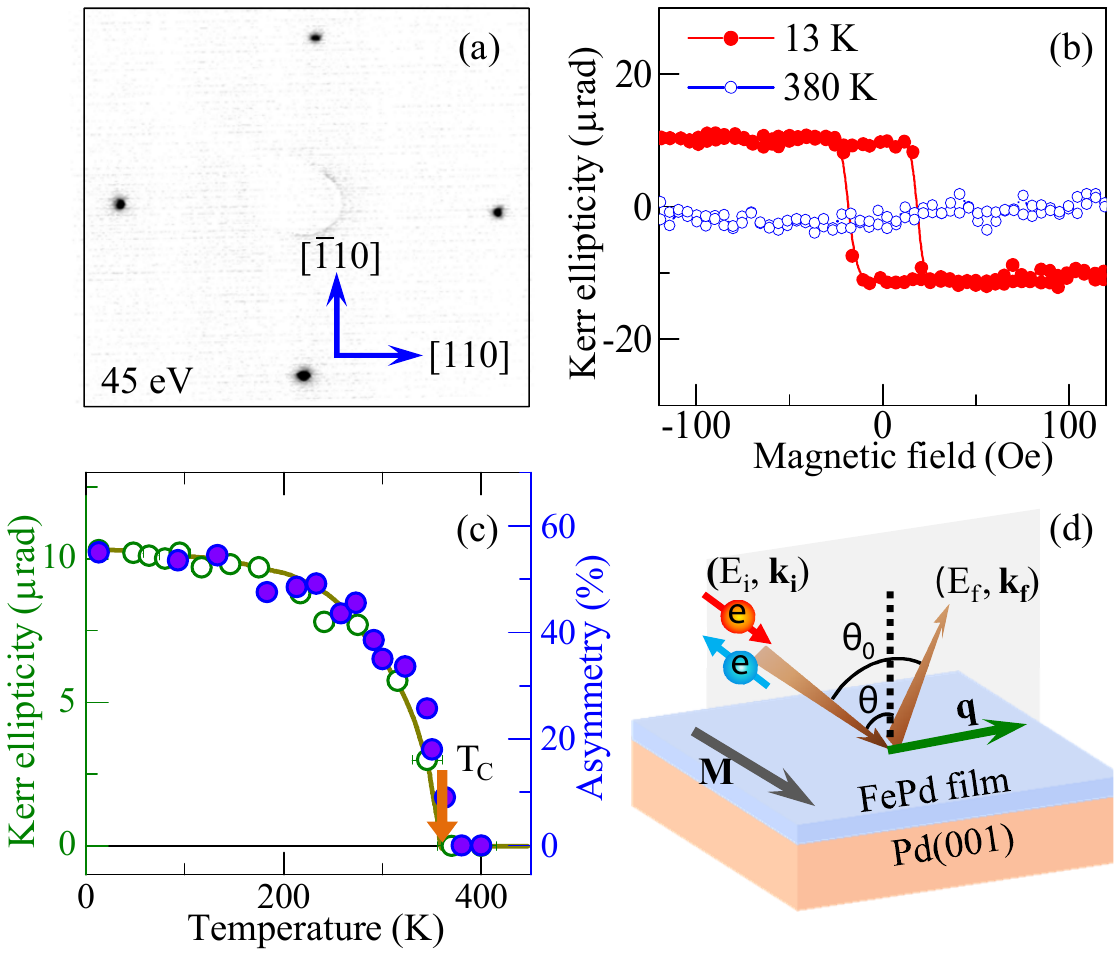}}
\caption{(color online) (a) The LEED pattern of the sample taken at a primary electron energy of 45 eV. (b) LMOKE hysteresis loops recorded at the temperatures of 13 and 380 K, labeled
by red solid and blue open circles, respectively. (c) The Kerr ellipticity, open circles, and the SPEELS spin asymmetry $A$, field circles, as a function of temperature.  (d) A schematic representation of the scattering geometry. }
\label{Fig1}
\end{figure}

One atomic layer of Fe was deposited on the clean Pd(001) surface and subsequently annealed at 400 K for 10 minutes in order to obtain an ultrathin FePd alloy film with the thickness of two atomic layers \cite{Meyerheim2006}. The sample quality was checked by low-energy electron diffraction (LEED) and Auger electron spectroscopy. The LEED pattern is presented in Fig. \ref{Fig1} (a),  showing a sharp 1 $\times$ 1  pattern with a well-defined surface structure. The magnetic hysteresis loop was measured by longitudinal magneto-optical Kerr effect (LMOKE) with the magnetic field applied along the [$1\bar{1}0$]-direction. The hysteresis loops recorded at $T=13$ and 380 K are presented in Fig. \ref {Fig1} (b). The film shows a rectangular hysteresis loop at low-temperatures, indicating a well-defined ferromagnetic state. Figure \ref {Fig1} (c) shows the Kerr ellipticity versus temperature. The Kerr ellipticity decreases with temperature and disappears at $T_C \simeq$ 380 K, indicating a ferromagnetic to paramagnetic phase transition.

Magnetic excitations were investigated by means of spin-polarized high resolution electron energy loss spectroscopy (SPEELS).  The technique is particularly suitable for investigation of terahertz magnons in ultrathin ferromagnetic films, due to its monolayer sensitivity and covering a wide range of wave vectors \cite{Vollmer2003, Tang2007, Qin2013, Zakeri2014}. A schematic representation of the scattering geometry used in our experiments is shown in Fig. \ref{Fig1} (d). The wave vector $\vec{q}$ of the excited magnons is given by $\vec{q} = \vec{k_{i}} - \vec{k_{f}}$, where $\vec{k_{i}}$ ($\vec{k_{f}}$) is the wave vector of the incident (scattered) electrons. In this geometry the magnon propagation direction (or $\vec{q}$) is perpendicular to the direction of the sample magnetization $\vec{M}$. The incident beam energy was $6$ eV. The measurements were performed along the $\bar{\rm\Gamma}$-$\bar{\rm X}$ direction. For all experiments presented here a longitudinally spin-polarized beam (polarization vector parallel/antiparallel to $\vec{M}$) was used. This allowed us to measure the spin asymmetry as a function of temperature and thereby monitor the evolution of the long-range magnetic order with temperature. Note that irrespective of the type of beam polarization, the excitations observed in SPEELS are the well-known collective spin excitations described by established spin wave theories \cite{Note1}.

\begin{figure*}[!htp]
\vspace{12pt} \center
\resizebox*{1.6\columnwidth}{!}{\includegraphics{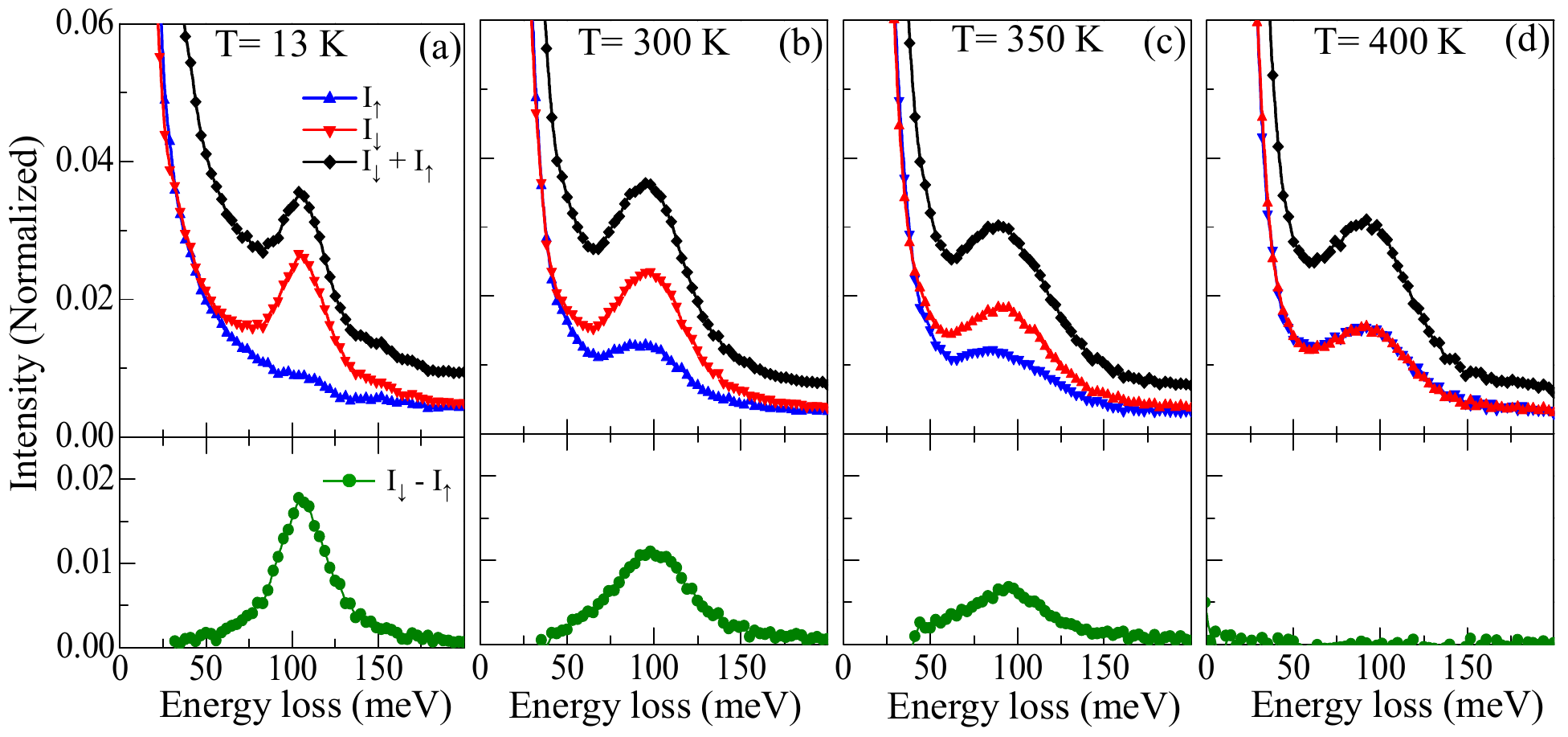}}
\caption{(color online) (a-d) SPEELS spectra recorded at $q=1.1$  \AA$^{-1}$ on a two atomic-layer
thick Fe$_{50}$Pd$_{50}$  film on Pd(001) at $T=13$ K (a),
300 K (b), 350 K (c) and 400 K (d) .  The red ($I_{\downarrow}$) and blue ($I_{\uparrow}$) spectra represent
the intensity of the scattered electrons for the spin polarization of
the incident beam parallel and antiparallel to the sample magnetization,
respectively. The difference ($I_{\downarrow}$-$I_{\uparrow}$) and total spectra ($I_{\downarrow}$+$I_{\uparrow}$) are represented by green and black solid symbols, respectively.}
\label{Fig2}
\end{figure*}


The SPEELS spectra were recorded at $q=0.6$, 0.8, and 1.1 \AA $^{-1}$. All spectra were recorded over a temperature range from 13  to 400 K. Figure \ref{Fig2} shows typical SPEELS spectra recorded at $q=1.1$ \AA$^{-1}$ (close to the $\bar{\rm{X}}$-point) and for four different temperatures. $I_{\downarrow}$ ($I_{\uparrow}$)
denotes the intensity of the scattered beam when the spin polarization of the incident electrons is parallel (antiparallel) to $\vec{M}$. In Fig. \ref{Fig2} (a) a well-defined peak appears at about 105 meV in the $I_{\downarrow}$
spectrum at T= 13 K. This peak is ascribed to the magnon excitation \cite{Zakeri2014}. Due to the conservation law of the total angular momentum, the magnons are only excited by incidence of minority electrons ($S=-\frac{1}{2}\hbar$, spin polarization parallel to $\vec{M}$).  The total spectrum, defined as $I_{\downarrow} + I_{\uparrow}$, is represented by solid diamonds and the difference spectrum, defined as $I_{\downarrow} - I_{\uparrow}$, is represented by solid circles. As the temperature increases the peak intensity in the $I_{\downarrow}$ spectrum gradually decreases. At the same time the peak intensity in the $I_{\uparrow}$ spectrum gradually increases. At the temperature of 400 K (1.05 $T_C$), the peak intensity in both spin spectra is exactly the same. Consequently, the intensity of the difference spectrum, shown in the lower part of Fig. \ref{Fig2}, decreases with temperature. At $T \geq T_C$ , the intensity of the difference spectrum becomes zero. The peak intensity in the total spectra ($I_{\downarrow} + I_{\uparrow}$) is nearly temperature independent. A similar behavior was observed for the spectra recorded at  $q=0.6$ and 0.8 \AA$^{-1}$.
Such a behavior of the SPEELS spectra as a function of temperature is expected and can be understood based on the following picture. At the lowest temperature ($T \longrightarrow 0$ K), the film is ferromagnetic. After magnetizing, the system is a single domain ferromagnet with a well-defined total magnetization $\vec{M}$. The direction of $\vec{M}$ is defined as the quantization axis. In the SPEELS experiments, due to the conservation law of the total angular momentum, magnons can only be excited when the incident electron's spin is of minority character (parallel to $\vec{M}$). Therefore, a pronounced magnon peak appears in the $I_{\downarrow}$  spectrum and no peak in $I_{\uparrow}$ \cite{Note1} . As the temperature increases, the thermal energy increases, leading to the formation of the magnetic domains. At temperatures below $T_C$, the majority of domains have still their local magnetization along the quantization axis and only a few of them have their magnetization pointing along any other arbitrary direction. Under this condition, a weak magnon peak appears also in the $I_{\uparrow}$  spectrum. As temperature increases the domain size becomes smaller and the number of domains having an arbitrary magnetization direction increases. This leads to an increase in the magnon peak intensity in the $I_{\uparrow}$  spectrum.  When the temperature reaches $T_C$, the peak intensity in both $I_{\uparrow}$ and $I_{\downarrow}$ becomes identical. At this temperature the thermal fluctuations become so large that the long-range magnetic order does not exist anymore and the sample undergoes a phase transition. This means that the asymmetry ($A=\frac{I_{\downarrow} - I_{\uparrow}}{I_{\downarrow} + I_{\uparrow}}$) of the magnon peak is proportional to $M$. Indeed there is a very nice correlation between the measured Kerr signal and $A$ [see Fig. \ref{Fig1}(c)]. The magnetization vanishes at $T_C$, where the long range magnetic order is absent. However, due to the existence of short-range magnetic order the magnons do exist even in the paramagnetic phase, in line with the theoretical prediction \cite{Razee2002,Antropov2005}. Figure \ref{Fig2} (d) is a direct evidence of the existence of terahertz magnons at and above $T_C$.
This observation can be understood based on the fact that terahertz magnons are governed by the short-range magnetic interactions. A short spin-spin correlation length is, in turn, sufficient to stabilize local ordering on atomic spins which are a few lattice sites away. Terahertz magnons can be excited on these length-scales \cite{Antropov2005}.

\begin{figure*}[!htb]
\vspace{12pt} \center
\resizebox*{1.99\columnwidth}{!}{\includegraphics{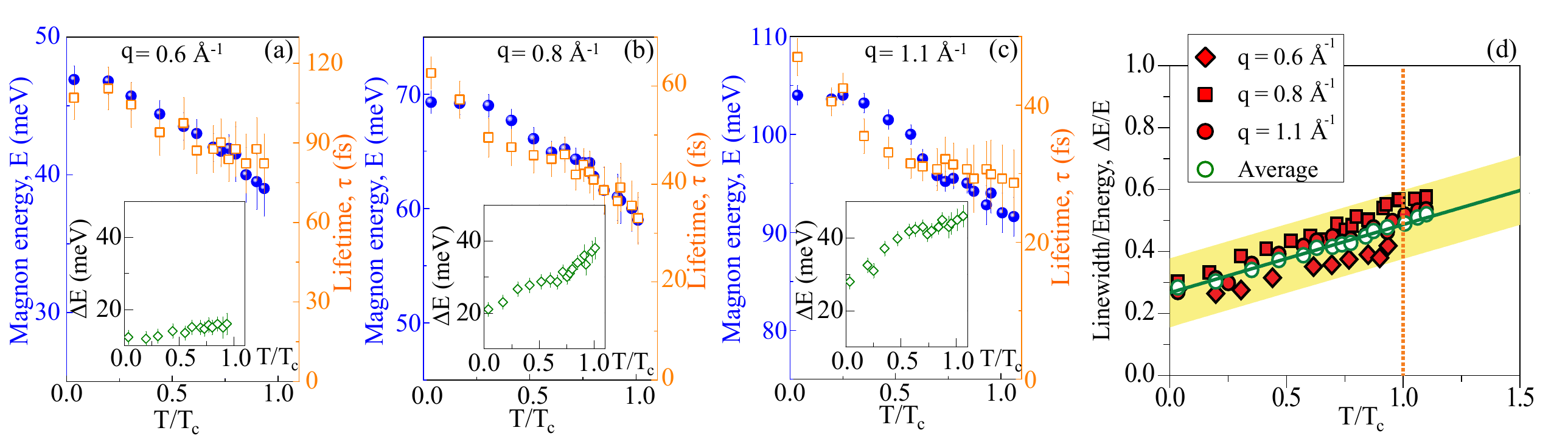}}
\caption{(color online)Magnon energy (solid circles) and lifetime
(open squares) as a function of the reduced temperature at $q= 0.6$ \AA$^{-1}$ (a), 0.8 \AA$^{-1}$ (b), and 1.1 \AA$^{-1}$ (c). The insets represent the corresponding linewidth versus the reduced temperature. (d) The temperature dependence of the decay parameter ($\alpha=\Delta E/E$). The field symbols denote the results of different wave vectors. The average is represented by open circles. The solid line indicates a linear fit to the data ($y=0.267+0.22x$).}
\label{Fig3}
\end{figure*}

The linewidth of spectra provides information on the magnon lifetime. In order to extract such information both the total ($I_{\downarrow}$+$I_{\uparrow}$) and difference spectra ($I_{\downarrow}$-$I_{\uparrow}$) are fitted using a convolution of a Gaussian and a Lorentzian function, in which the Gaussian represents the instrumental broadening and the Lorentzian represents the intrinsic lifetime broadening \cite{Zakeri2012, Qin2013}.
The analysis of both total and difference spectra leads to the same results. Hence in the following we discuss the results of analysis of the total spectra, as they can provide the information above $T_C$. Figure \ref{Fig3} shows the magnon energy $E$ and lifetime $\tau$ versus the reduced temperature.The temperature dependence of the linewidth $\Delta E$  is provided in the insets. The lifetime is obtained using the expression: $\tau=2\hbar/\Delta E$. The magnon energy decreases as the temperature increases. For instance,
at $q$ = 0.6 \AA$^{-1}$, the energy disperses from 47 meV to 39 meV i.e.,
a decrease of about 15\%. Similarly at $q$ = 1.1 \AA$^{-1}$  the energy reduces from
104 meV to 92 meV (about 15\%). Since the energy of high-energy magnons is entirely determined by the magnetic exchange interaction, the renormalization of exchange interaction at elevated temperatures results in a reduction of magnons' energy. Generally, with increasing temperature the exchange interaction decreases \cite{Dietrich1976,Szilva2013,Zakeri2014a}. This fact leads to the reduction of the magnons' energy. Analysis based on a simple nearest neighbor Heisenberg model indicates that the effective exchange interaction decreases by about 15\% when increasing the temperature from 0.03 $T_C$  to $T_C$. This is in line with the recent spin dynamic simulations, where a magnon softening is predicted for Fe films as a result of temperature \cite{Bergman2010,Bergqvist2013, Etz2015, Rodrigues2016}. The lifetime also reduces as temperature increases. At $q$ = 0.6 \AA$^{-1}$, it decreases from 110 femtoseconds (fs) to 80 fs with the temperature increasing from 13 K to $T_C$. At $q$ = 1.1 \AA$^{-1}$, it decreases from 50 fs to 29 fs.

At low temperatures the main mechanism leading to the damping of terahertz magnons in itinerant ferromagnets is their decay into the single-particle Stoner excitations (known as Landau damping) \cite{Buczek2011a,Zhang2012,Zakeri2013,Qin2015}.
Since single-particle excitations correspond to electronic transitions from majority to minority states across the Fermi level, the density of Stoner states is usually given by the joint probability of creating a hole in majority states and an electron in minority states. The theory which considers both magnons and Stoner excitations on an equal footing can perfectly explain the magnon lifetime in this material at low temperatures \cite{Buczek2011a,Qin2015}. In our theory we included the temperature induced renormalization of the electronic structures and found out that the Landau damping of magnons is nearly unchanged. This is due to the fact that the renormalization of the electronic structures as a function of temperature is very small. Hence, one can exclude a strong temperature dependence of the Landau damping. At higher temperatures, other damping mechanisms become active. Therefore the reduction of the magnons' lifetime at higher temperatures is due to the presence of other decay channels e.g. multi-magnon scattering. The multi-magnon scattering damping is usually referred to as a damping mechanism of a well defined magnon mode as a result of its scattering with/into other magnon modes. This mechanism has already been well investigated in the case of small wave vector excitations, mainly ferromagnetic resonance mode with $q=0$ (see for example \cite{Zakeri2007} and references therein). However, it has not been investigated for the case of terahertz excitations and as a function of temperature. We provide, for the first time, quantitative experimental data of this damping parameter as a function of temperature [see Fig. \ref{Fig3} (d)].  Our experiment shows that this damping mechanism becomes more important at higher temperatures. This is expected due to two reasons. Firstly, the number and the type of thermally excited magnons increase with temperature. Secondly, the scattering at the domain boundaries becomes more important, because the number of ferromagnetic domains increases with temperature.  Our results indicate that the damping induced by this mechanism increases monotonously with temperature and can even become comparable to the intrinsic Landau damping of the system.

In order to better see the temperature effects on both the magnon energy and lifetime and also on their real space propagation, we introduce a dimensionless decay parameter $\alpha=\Delta E/E $. This parameter shall also describe the character of magnons at a given temperature. Let us consider two extreme cases. If the linewidth broadening is much smaller than the magnon energy (the limit of small $\alpha$), the magnons shall appear as well-defined excitations with a long lifetime, which behave as propagating quasiparticles. For the case in which the linewidth broadening is comparable to the energy ($\alpha \simeq 1$), the magnons do not exhibit a propagating character. Figure \ref{Fig3} (d) shows the temperature dependence of $\alpha$ for the magnons with  $q=0.6$, 0.8 and 1.1 \AA$^{-1}$. Our results indicate that $\alpha$  exhibits a rather moderate temperature dependence for all wave vectors. Extrapolation of the data to $T=0$ K results in a value of $\alpha (T\longrightarrow0) \simeq 0.27\pm0.04$. This is in excellent agreement with the results of our dynamical calculations, suggesting a value of about $\alpha (T=0) \simeq 0.28$ \cite{Qin2015}. As temperature increases, $\alpha$ monotonously increases, up to and beyond $T_C$. The monotonous increase of $\alpha$ has its origin in increasing (i) the population of different kinds of thermally excited magnons and (ii) the number of scattering events at the domain boundaries, as the number of domains increases with temperature.  At $T_C$, $\alpha$ reaches the value of about 0.5. This observation clearly indicates that the high energy magnons sustain their propagating character at $T_C$. Extrapolation of the data to higher temperatures using the most simplest function i.e. a linear fit implies that $\alpha$ reaches the unity at temperatures far above $T_C$  (at about $3T_C$). Another way to confirm that terahertz magnons sustain their propagating nature above $T_C$ is to visualize their dynamics by constructing the magnon wave packets in real space. Our analysis for a magnon wave packet with $q=0.8$ \AA$^{-1}$ showed that this wave packet can propagate about 2.9 nm before its amplitude is decayed to 1\% of its initial value \cite{Note2}.

In summary, we discussed the behavior of magnons as a function of temperature and across $T_C$. We provided experimental evidence that terahertz magnons can be efficiently excited above $T_c$. The magnons' energy and lifetime decrease with temperature in a rather complex manner, which strongly depends on the wave vector. The high-energy magnons sustain their propagating character even at temperatures far above $T_C$. We provide, quantitatively, the damping parameter of terahertz magnons as a function of temperature and discuss the possible damping mechanisms. In addition to the fact that our results are important for developing the theory of spin excitations at a finite temperature, our finding is of particular importance for the research in the field of magnonics. As terahertz magnons can be very efficiently excited in the paramagnetic state of a ferromagnet, they can be used for information transfer/processing. In other words one can use the terahertz magnons, excited in a device made of a ferromagnetic film, for data processing without being worried about the effects caused by the magnetic phase transition. Furthermore, our results indicate that the effects  associated with the collective spin excitations can also survive above $T_C$. For instance the spin-Seebeck effect can eventually exist also in the paramagnetic state, where the long range magnetic order is absent (spin caloritronics is not restricted to the ferromagnetic materials), confirming the recent experimental observation of this effect in the paramagnetic state \cite{Wu2015}. Our results shall bring the topics of ``paramagnonics" and ``spin caloritronics using paramagnets" into discussion.

The work has partially been supported by the Deutsche Forschungsgemeinschaft (DFG) through the Heisenberg Programme ZA 902/3-1 and the DFG grant ZA 902/4-1.
\bibliographystyle {apsrev}
\bibliography {./Qin}

\end{document}